\documentclass[traditabstract]{aa}
\usepackage{graphicx}
\usepackage{txfonts}
\usepackage{booktabs}

\newcommand{\Rlag}{R_{\rm \tau_K}}

\newcommand{\Rring}{R_{\rm ring}}

\newcommand{\Vsys}{V^2_{\rm sys}}

\usepackage{natbib}
\begin{document}

\title{The innermost dusty structure in active galactic nuclei as
  probed by the Keck interferometer}


\author{Makoto Kishimoto$^1$,
       Sebastian F. H\"onig$^2$,
       Robert Antonucci$^2$, 
       Richard Barvainis$^3$, 
       Takayuki Kotani$^4$, 
       Konrad R.~W.~Tristram$^1$,
       Gerd Weigelt$^1$, 
       \and
       Ken Levin$^5$
       }

\institute{
  $^1$Max-Planck-Institut f\"ur Radioastronomie, Auf dem H\"ugel 69,
  53121 Bonn, Germany; \email{mk@mpifr-bonn.mpg.de}.  
  $^2$Physics Department, University of California, 
  Santa Barbara, CA 93106, USA. $^3$National
  Science Foundation, 4301 Wilson Boulevard, Arlington, 
  VA 22230, USA.  
  $^4$ISAS, JAXA, 3-1-1 Yoshinodai, Sagamihara, Kanagawa, 
  229-8510 Japan. 
  $^5$Silver Spring Observatory, 10712 Meadowhill Rd, 
Silver Spring, MD 20901
          }

\date{Submitted 3 November 2010 / Accepted 23 December 2010}

\authorrunning{Kishimoto et al.}

\titlerunning{The innermost dusty structure in active galactic nuclei}

 
\abstract{ 

  We are now exploring the inner region of Type 1 active galactic
  nuclei (AGNs) with the Keck interferometer in the
  near-infrared. Adding to the four targets previously studied, we
  report measurements of the K-band (2.2~$\mu$m) visibilities for four
  more targets, namely AKN120, IC4329A, Mrk6, and the radio-loud QSO
  3C273 at $z$=0.158.  The observed visibilities are quite high for
  all the targets, which we interpret as an indication of the partial
  resolution of the dust sublimation region.  The effective ring radii
  derived from the observed visibilities scale approximately with
  $L^{1/2}$, where $L$ is the AGN luminosity.  Comparing the radii
  with those from independent optical-infrared reverberation
  measurements, these data support our previous claim that the
  interferometric ring radius is either roughly equal to or slightly
  larger than the reverberation radius.  We interpret the ratio of
  these two radii for a given $L$ as an approximate probe of the
  radial distribution of the inner accreting material.  We show
  tentative evidence that this inner radial structure might be closely
  related to the radio-loudness of the central engine.  Finally, we
  re-observed the brightest Seyfert 1 galaxy NGC4151. Its marginally
  higher visibility at a shorter projected baseline, compared to our
  previous measurements obtained one year before, further supports the
  partial resolution of the inner structure. We did not detect any
  significant change in the implied emission size when the K-band flux
  was brightened by a factor of 1.5 over a time interval of one year.}


\maketitle
%

\section{Introduction}

With long-baseline interferometry in the mid-IR, the inner dust
distribution in active galactic nuclei (AGN) has been studied over the
past several years (e.g. early results on Seyfert~2 galaxies by
\citealt{Jaffe04}, \citealt{Tristram07}; radio galaxy, \citealt
{Meisenheimer07}; Type~1 AGNs, \citealt{Beckert08}, \citealt
{Kishimoto09}, \citealt{Burtscher09}; snapshot survey, \citealt
{Tristram09}).  In the near-IR, after the very early results for the
brightest Seyfert~1 galaxy NGC4151 \citep{Swain03} and Seyfert~2
galaxy NGC1068 \citep{Wittkowski04}, long-baseline interferometry has
now started to explore the innermost dusty structure extensively at
milli-arcsec (mas) resolution (\citealt{Kishimoto09KI}, hereafter
Paper~I; \citealt{Pott10}).  We focus on Type 1 AGNs, which are
considered to provide us with a direct view of the region and thus can
be used to scrutinize the inner structure.  Here we report on new
measurements for four more targets, which double the sample size of
our previous study in Paper~I.

\begin{figure*}
\centering \includegraphics[width=19cm]{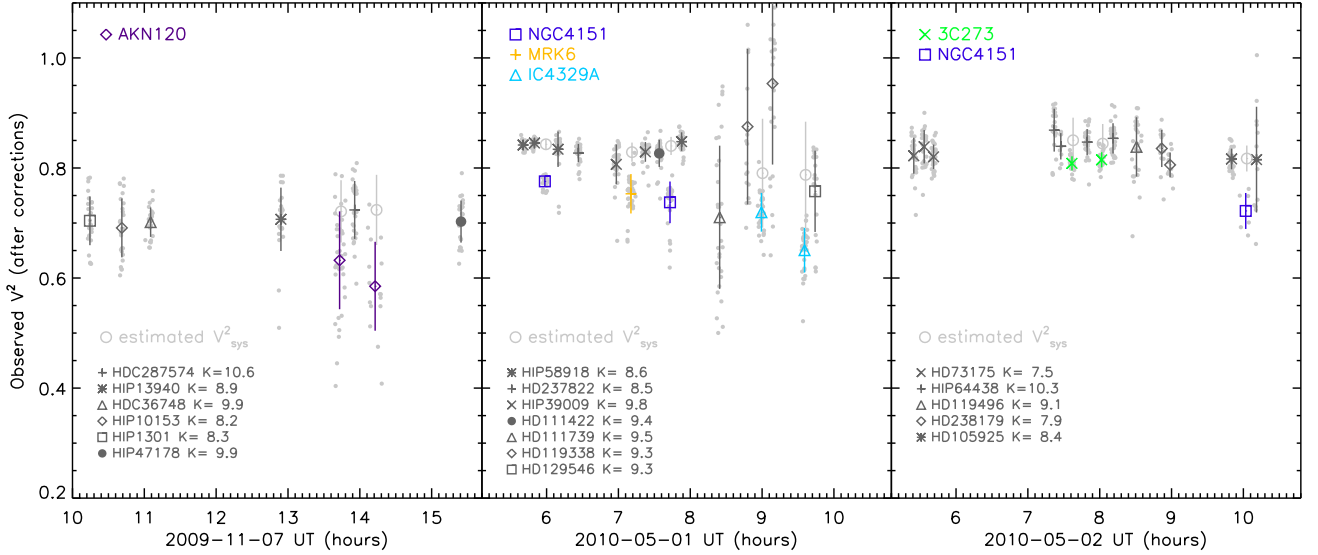}
\caption{\small Observed $V^2$ (after flux-ratio and flux-bias corrections;
  see sect.\ref{sec_obs_ki}) plotted against observing time. Gray dots
  are individual measurements for blocks of 5 sec each.  These blocks
  are averaged into scans over a few to several minutes each.  These
  scans are indicated by various symbols with error bars corresponding
  to statistical errors, calculated as standard deviations of the
  blocks within each scan.  Gray circles are the estimated system
  $V^2$ at the time of target observations (plotted slightly shifted
  in time for clarity).}
\label{fig_rawvis}
\end{figure*}

\begin{table*}

\caption[]{Summary of the results of our KI observations and radio
  data for the present and remaining previous targets.}
{\tiny
\begin{tabular}{lccccccccccccccc}
\hline
name  & date (UT) & $z^a$  & scale$^b$ & $E_{B-V}$$^c$ & $B_{p}$$^d$ & PA & $V^2$ & \multicolumn{2}{c}{$\Rring$$^e$}  & \multicolumn{2}{c}{radio flux$^f$}  & $A_{V}$ \\ 
\cmidrule(rl){9-10}\cmidrule(rl){11-12}
     &       & corr.& (pc mas$^{-1}$)  &  (mag)  & (m)   & (\degr)   &           & (mas)        & (pc)        & (mJy) & (GHz) & (mag)   \\ 
\hline
AKN120   & 2009-11-07 & 0.0327 & 0.653 & 0.128 & 74.6 & 31.9 & 0.843$\pm$0.097 & 0.56$\pm$0.19 & 0.36$\pm$0.12 &  9.3$^g$ & 1.4 \\
IC4329A  & 2010-05-01 & 0.0170 & 0.345 & 0.059 & 69.8 & 49.6 & 0.868$\pm$0.084 & 0.54$\pm$0.18 & 0.19$\pm$0.06 & 10.7$^h$ & 8.4 & 2$\pm$1$^n$ \\
MRK6     & 2010-05-01 & 0.0188 & 0.382 & 0.136 & 71.5 &$-$1.4& 0.909$\pm$0.048 & 0.44$\pm$0.12 & 0.17$\pm$0.05 & 39.5$^i$ & 8.4 \\
3C273    & 2010-05-02 & 0.159  &  2.75 & 0.021 & 83.5 & 39.9 & 0.958$\pm$0.034 & 0.25$\pm$0.10 & 0.69$\pm$0.28 & 3.20E4$^j$&4.8 \\
NGC4151  & 2010-05-01 & 0.00414& 0.0855& 0.028 & 72.4 & 51.0 & 0.920$\pm$0.017 & 0.40$\pm$0.05 &0.034$\pm$0.004& 125.$^k$ & 4.9 \\
         & 2010-05-01 &         &       &       & 81.3 & 40.7 & 0.878$\pm$0.049 & 0.45$\pm$0.10 &0.038$\pm$0.008& \\ 
         & 2010-05-02 &         &       &       & 84.9 & 21.7 & 0.883$\pm$0.048 & 0.42$\pm$0.09 &0.036$\pm$0.008&                       & \\ 
\hline
NGC4051     &         & 0.00309 & 0.0639 &       &      &      &                 &               &               & 6.0$^k$ & 4.9 \\
MRK231      &         & 0.0427  & 0.841  &       &      &      &                 &               &               & 173.$^l$& 5.0 \\
IRAS13349$^o$  &                       & 0.109   & 1.98 & & &      &                 &               &               & 19.9$^m$& 1.4\\
\hline
\end{tabular}
\\
$^a$ CMB corrected value from NED. 
$^b$ $H_0=70$ km s$^{-1}$ Mpc$^{-1}$, $\Omega_{\rm m}=0.3$, and $\Omega_{\Lambda}=0.7$.
$^c$ Galactic reddening from \cite{Schlegel98}. $^d$ Projected baseline lengths. 
$^e$ Thin-ring radius.
$^f$ Radio flux observed at frequency given. 
For radio-quiet objects, flux from linear (double or triple) structure at $\sim$100~pc scale is given, except for 
AKN120, IC4329A, and IRAS13349+2438, for which flux from a source unresolved at resolution of $6"$
(3.9kpc), $1"$ (350pc), and $5.4"$ (11kpc), respectively, is given.
$^g$ \cite{Condon98}.
$^h$ \cite{Nagar99}.
$^i$ \cite{Schmitt01}.
$^j$ \cite{Tingay03}.
$^k$ \cite{Ulvestad84}.
$^l$ \cite{Ulvestad99}.
$^m$ \cite{White97}.
$^n$ \cite{Winkler92I}. 
$^o$ IRAS13349+2438.
}

\label{tab_obs}
\end{table*}

\section{Observations and data reduction}\label{sec_obs}

\subsection{Keck interferometry}\label{sec_obs_ki}

The Keck Interferometer (KI; \citealt{Colavita03}) combines the two 
beams from the two Keck 10~m telescopes which are separated by 85~m 
along the direction 38\degr\ east of north. We observed Type 1 AGNs 
listed in Table~\ref{tab_obs} and associated calibrators with the KI in 
Nov 2009 and May 2010.  The targets were partly chosen based on the 
bright optical magnitudes measured from the pre-imaging data obtained 
in Oct 2009 at Tiki Observatory (French Polynesia) by N.~Teamo and
J.~C.~Pelle, and at Silver Spring Observatory (USA). 

The Adaptive Optics system at each Keck telescope was locked on the
nucleus of each target galaxy at visible wavelengths. The fringe
tracker was operated in the K-band at the rates of 100Hz and 200Hz in
our observing runs in Nov 2009 and May 2010, respectively. The
angle-tracking was performed in H-band at the rate of either 40 or
80~Hz. The data were first reduced with {\sf Kvis}\footnotemark[1] to
produce raw squared visibility ($V^2$) data averaged over blocks of
5sec each.

During the observing night in Nov 2009, we often detected a quite
uneven level of flux from the two telescopes, resulting in visibility
attenuation.  In the May 2010 run, this was seen only for the
calibrators observed at large air masses.  The standard software {\sf
  wbCalib}\footnotemark[2] has been shown to be able to correct for
this effect\footnotemark[3], but we also rejected blocks associated
with a discrepantly large flux ratio of the two beams
($\gtrsim$8). The wide-band side of the data were then reduced using
{\sf wbCalib} with the correction for the flux ratio and flux bias.
The blocks were averaged into scans over a few to several minutes
each, with its error being estimated as a standard deviation within a
scan.  Figure~\ref{fig_rawvis} shows the observed visibilities of all
the targets and calibrators (after the corrections above), plotted
against the observing time.

\footnotetext[1]{\tiny{http://nexsci.caltech.edu/software/KISupport/v2/V2reductionGuide.pdf}}

\footnotetext[2]{\tiny{http://nexsci.caltech.edu/software/V2calib/wbCalib/index.html}}

\footnotetext[3]{\tiny{http://nexsci.caltech.edu/software/KISupport/dataMemos/KI\_ratio.pdf}}

All the calibrators used\footnotemark[4] (Fig.~\ref{fig_rawvis}) are
expected to be unresolved by the KI at K-band ($V^2 \ge 0.999$).  The
calibrators span a relatively wide range of brightness, and some were
found to have lower injected flux counts at K-band than the science
targets (for the same K-band magnitude, calibrator counts are often
higher due partly to their bluer colors which lead to a better AO
performance in the optical.)  As in Paper~I, we assign 0.01 to be the
possible systematic uncertainty in the system visibility estimate,
based on the difference between the corrected visibilities of brighter
calibrators and those of fainter ones, though in many cases the total
uncertainty is dominated by much larger statistical errors.  For each
target measurement, the system visibility $\Vsys$ was estimated from
these calibrator observations with {\sf wbCalib} using its time and
sky proximity weighting scheme (using the default\footnotemark[5] time
constant of 1 hour and distance of 15$^{\circ}$), yielding the $\Vsys$
measurements indicated by the gray circles in
Fig.~\ref{fig_rawvis}. Averages were taken for those objects with two
calibrated measurements adjacent in time that were consistent with
each other (Fig.~\ref{fig_rawvis}; AKN120, IC4329A, 3C273).  The final
calibrated visibilities with the errors including the systematic
uncertainty added in quadrature are shown in Table~\ref{tab_obs}.

\footnotetext[4]{The star HD240991 was excluded from the calibration,
  since two observations showed systematically lower visibilities than
  other calibrators.}

\footnotetext[5]{\tiny{http://nexsci.caltech.edu/software/V2calib/wbCalib/index.html}}

\subsection{UKIRT imaging}

\begin{figure}
\centering \includegraphics[width=8cm]{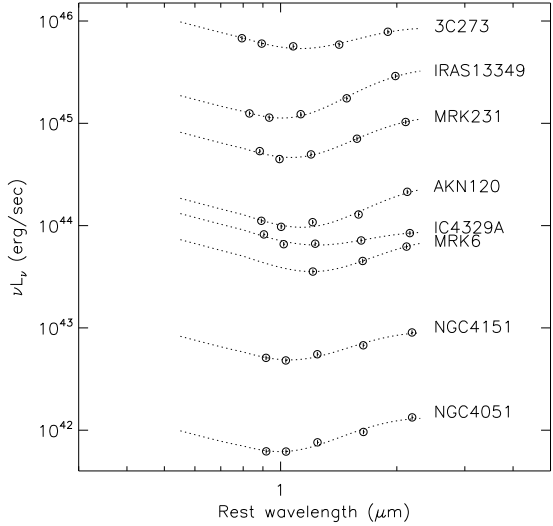}
\caption{\small Flux of the nuclear PSF component in WFCAM images (except for
  Mrk6, for which we used 2MASS images) derived from 2D fits. The SEDs
  fitted with a black body and a power-law $f_{\nu} \propto
  \nu^{+1/3}$ (see sect.~\ref{sec_vis}) are shown by dotted lines.}
\label{fig_sed}
\end{figure}

To obtain contemporaneous flux measurements of the variable nuclei, we
observed all the KI targets with WFCAM on UKIRT (except for Mrk~6,
which was outside the UKIRT pointing limit) about five to six weeks
after each KI observing run, which is generally shorter than the
variability timescale, at five broad bands from 0.9 to 2.2 $\mu$m.
The observations were done under the UKIRT service program. The data
were handled in the same way as described in Paper~I. The wide FOV of
WFCAM provided simultaneous PSF measurements from stars in the same
field.  These PSFs were used to decompose each image into the nuclear
PSF component and host galaxy, leading to the SED measurement of the
nuclear point source in each AGN.  The measured fluxes are summarized
in Table~\ref{tab_corr}.  The SEDs, after correcting for reddening in
both our Galaxy and for IC4329A in the host galaxy
(Table~\ref{tab_corr}), are shown in Fig.~\ref{fig_sed}, together with
the four SEDs from Paper~I.

\section{Results}

\subsection{Visibilities and ring radii}\label{sec_vis}

For all of our targets, quite high visibilities were observed
(Table~\ref{tab_obs}), which was also the case for the previous four
targets in Paper~I.  We interpret these high visibilities as an
indication of the partial resolution of the dust sublimation region as
we argued in Paper~I and discussed in sects.~\ref{sec_ngc4151} and
\ref{sec_disc} below.  The most simplistic, physically-motivated
representation of the dust sublimation region seen face-on would be a
ring geometry.  To measure its effective radius regardless of the
structure details (e.g.  smooth versus clumpy distribution
[\citealt{Krolik88}], this information probably not being contained in
the visibilities at low spatial frequencies well before the first
null), we deduce in Table~\ref{tab_obs} a thin-ring radius $\Rring$
that corresponds to the observed visibility at the projected baseline
for each target observation.

The fractional uncertainty in the ring radius $\sigma_{\Rring}/\Rring$
for the data at the low spatial frequencies well before the first null
can be estimated in the same way as in the Gaussian geometry case, 
given as
$$
  \frac{\sigma_{\Rring}}{\Rring} = \frac{1}{-2 \ln V^2}\frac{\sigma_{V^2}}{V^2}, 
$$
where $\sigma_{V^2}$ is the total uncertainty in $V^2$.  When we have
multiple data at different baselines as in the case of NGC4151, we
estimate the overall fractional uncertainty in the same way by
replacing $V^2$ and $\sigma_{V^2}/V^2$ with their mean over the data
points.

\begin{figure*}
\centering 
\includegraphics[width=10.1cm]{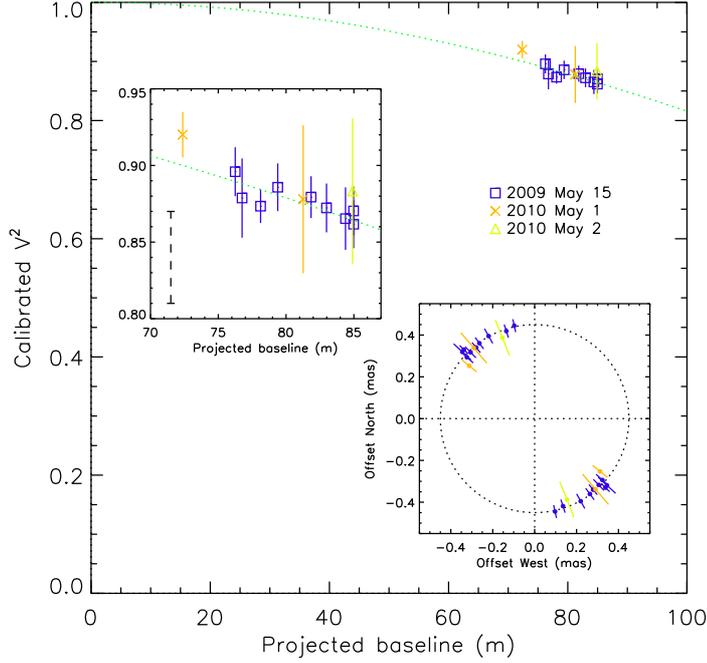}
\caption{\small Calibrated $V^2$ for NGC4151 as a function of projected
  baselines. Squares show our previous data taken in May 2009
  (Paper~I), while crosses and triangle are the new data taken on 2010
  May 1 and 2, respectively, all being enlarged in the left inset.
  The absolute $V^2$ calibration is said to have a possible systematic
  error up to 0.03 (see sect.~\ref{sec_ngc4151}), which is shown as a
  dashed vertical bar in this inset.  The dotted line in each panel
  shows the best-fit visibility curve of a thin-ring model with a
  radius of 0.45 mas for our previous May 2009 data.  In the right
  inset, ring radii corresponding to each data point are plotted along
  the PA of each projected baseline. Note that the correction for the
  accretion disk and host galaxy contributions is not incorporated in
  this figure, and does not significantly change the ring radius (see
  Table~\ref{tab_corr}).}
\label{fig_ngc4151}
\end{figure*}

We note that the exact inclination of the structure is not known, 
thus the resulting possible elongation of the ring due to inclination
is still generally unknown. However, we note that we do not
see a significant PA dependence in the effective ring radii in NGC4151
over the PA range from $\sim$10$\degr$ to $\sim$50$\degr$ (Paper~I;
see also Fig.~\ref{fig_ngc4151} described below), where the inner
system axis is supposed to be at PA $\sim$90$\degr$ (optical
polarization, \citealt{Martel98}; radio jet, \citealt{Mundell03}), 
despite the object often being thought to be relatively
inclined at the narrow-line region scale (e.g. \citealt{Das05} and
references therein).

Since the visibility measurements are affected by the host galaxy flux
fraction within the FOV of the KI, and also by the flux contribution
from the unresolved accretion disk (AD), we estimated these two
quantities using our UKIRT imaging data in the same way as described
in sect.3 of Paper~I.  For Mrk6, we used the 2MASS images.  In brief,
we fitted each SED with a power-law form $f_{\nu} \propto \nu^{+1/3}$
for the AD (supported by both theory and observations;
\citealt{Kishimoto08}) and a black-body form for the dust (the
best-fit temperatures were $\sim$1200-1500K).  The results are stated
in Table~\ref{tab_corr} along with the ring radii corrected for these
effects. We note that the effects are generally very small, and the
two effects also tend to partially cancel each other.

\subsection{NGC4151 re-observation}\label{sec_ngc4151}

We obtained quite extensive data for NGC4151 in May 2009 (Paper~I),
but re-observed the target in May 2010 to measure the visibility
essentially at the shortest possible projected baseline under the
current delay line restriction in the KI, and to check for visibility
variability.  \cite{Pott10} compared their data on NGC4151 with that
of \cite{Swain03} and our 2009 data, and reported that the object has
not shown any significant variability in the K-band visibility.

Figure~\ref{fig_ngc4151} shows our re-observations plotted over the
May 2009 measurements.  The previous data (squares) showed a marginal
$V^2$ decrease over increasing baselines.  This is now apparently
supported by the higher $V^2$ observed at the shorter projected
baseline of $\sim$72~m.  We caution that, when we compare the data
from different nights, we should take into account the possible
systematic uncertainty in the absolute $V^2$ calibrations, while each
$V^2$ data point in Fig.~\ref{fig_ngc4151} is shown with only a
statistical error, which is adequate for comparing data within a given
stable night.  The systematic uncertainty could be up to $\sim$0.03
(footnote 6; indicated as a dashed vertical bar in the left inset of
Fig.~\ref{fig_ngc4151}), although this number is probably conservative
for the wide-band side data that we use here. This was deduced from
the systematic difference in $V^2$ between its observed and predicted
values for known-orbit binaries where the difference was comparable to
or smaller than statistical errors.  (From our data alone, we estimate
the systematic uncertainty of $\sim$0.01 after the flux-bias
correction, and this has been incorporated in the errors shown in
Table~\ref{tab_obs}, as described in sect.~\ref{sec_obs_ki}.)

\footnotetext[6]{\tiny{http://nexsci.caltech.edu/software/KISupport/dataMemos/kvis\_params.pdf}}

Having these in mind, we calculated the Pearson's linear correlation
coefficient between the projected baseline and calibrated $V^2$, by
weighting the May 2010 data conservatively using the uncertainty of
0.03 (when the statistical error is smaller than this value) while
keeping the statistical uncertainty for the 2009 data as weights. The
coefficient is $-0.766$, meaning in this case of 12 data points that
the confidence level for rejecting the null correlation is 99.6\%, or
2.9$\sigma$. (Note that weighting the 2009 data using the uncertainty
of 0.03 as well would in this case lead to even a larger
significance.)

The ring radius implied by the new measurements is consistent with the
radius deduced from the data in May 2009 within the errors even when
excluding the systematic uncertainty.  Thus, we did not detect any
significant variability in the interferometric size. This is
consistent with the result by \cite{Pott10}.  On the basis of our
quasi-simultaneous UKIRT SED measurements, the nuclear K-band flux has
brightened by a factor of 1.5 over the one year interval. On this
timescale, the overall inner radial dust distribution does not seem to
have changed significantly.  This conclusion remains the same even
after the host galaxy and accretion disk flux fractions are taken into
account.

\begin{figure}
\centering \includegraphics[width=9cm]{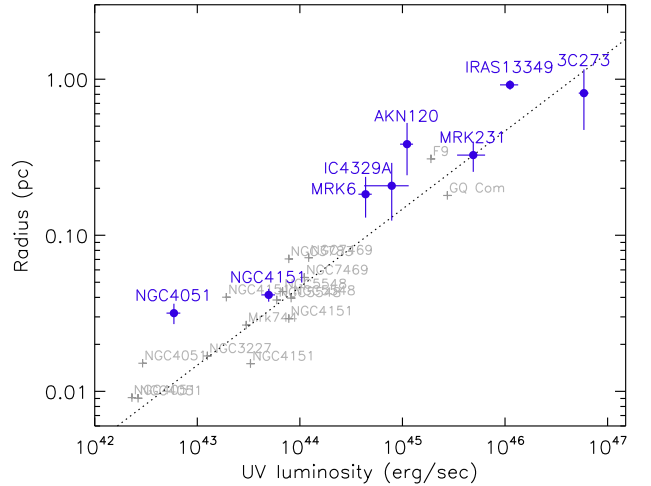}
\caption{\small Ring radius derived for each KI target (filled circles) after
  the correction for host galaxy and unresolved AD component, plotted
  against UV luminosity, or scaled V-band luminosity (extrapolated
  from the WFCAM SED; see sect.~\ref{sec_disc_lum}).  The
  reverberation radii against the same scaled V-band luminosity
  \citep[][and references therein]{Suganuma06} are also indicated by
  gray plus signs in addition to their fit (dotted line).}
\label{fig_radius_pc}
\end{figure}

\begin{table*}
\caption[]{Point-source flux from UKIRT imaging data and the KI results corrected for host galaxy and unresolved AD component.} 
{\tiny
\begin{tabular}{llcccccccccccccccccccc}
\hline
name    & date & \multicolumn{5}{c}{flux (mJy)$^a$} & \multicolumn{2}{c}{magnitude} & host$^b$  & AD    & $V^2$ corrected & $\Rring$ corr.   & $\Rlag$ fit$^d$ \\ 
\cmidrule(rl){3-7} \cmidrule(rl){8-9}
        & (UT) & Z & Y & J & H & K              & J & K                     & (\%)     & fraction$^c$ &                 & (pc)          & (pc)  \\ 
\hline
AKN120  & 2009-12-13 & 11.7 & 11.8 & 16.4 & 26.6 & 60.7  & 12.4 & 10.1 & 0.6$\pm$0.1 & 0.17$\pm$0.05 & 0.822$\pm$0.117    & 0.38$\pm$0.14     & 0.15  \\
IC4329A & 2010-06-13 & 15.0 & 16.0 & 23.9 & 41.6 & 75.5  & 12.0 &  9.8 & 0.9$\pm$0.2 & 0.29$\pm$0.09 & 0.839$\pm$0.119    & 0.21$\pm$0.08     & 0.13  \\
Mrk6$^e$& 1999-02-27 & ---  & ---  & 16.3 & 29.1 & 53.6  & 12.4 & 10.2 & 0.2$\pm$0.1 & 0.21$\pm$0.05 & 0.890$\pm$0.061    & 0.18$\pm$0.05     & 0.097 \\
3C273   & 2010-06-14 & 29.0 & 29.1 & 33.2 & 45.7 & 81.9  & 11.7 &  9.7 & $<$0.1      & 0.28$\pm$0.09 & 0.942$\pm$0.047    & 0.81$\pm$0.34     & 1.1   \\
NGC4151 & 2010-06-14 & 51.1 & 54.0 & 81.8 & 136. & 264.  & 10.7 &  8.5 & 0.2$\pm$0.1 & 0.14$\pm$0.05 & 0.912$\pm$0.021$^f$&0.037$\pm$0.007$^g$& 0.037 \\ 
\hline
\end{tabular}
\\
$^a$ Flux measurement uncertainty is $\sim$5 \% (see sect.3 in Paper I).
$^b$ Host galaxy flux fraction at K-band estimated for the KI's 50 mas
FOV in the AO-corrected images. $^c$ AD flux fraction of the point source at K-band.
$^d$ $\Rlag$ from UV luminosity using the fit by \cite{Suganuma06}.
$^e$ From 2MASS images.
$^f$ Value for the shortest baseline data. $^g$ Best-fit value for 2010-05-01 data.
}
\label{tab_corr}
\end{table*}

\section{Discussions}\label{sec_disc}

\subsection{Radius vs luminosity}\label{sec_disc_lum}

Figure~\ref{fig_radius_pc} shows the ring radii $\Rring$ derived from
the visibility, after the correction for the host galaxy and
unresolved AD component, as a function of the UV luminosity $L$,
defined here as a scaled optical luminosity of 6$\nu L_{\nu}$(5500\AA)
\citep{Kishimoto07}.  The flux at 5500\AA\ is here extrapolated from
the fitted SED at 0.8 $\mu$m by assuming the AD spectral shape of
$f_{\nu} \propto \nu^{0}$ (based on spectral index studies summarized
in Fig.~2 of \citealt{Kishimoto08}) and shown in Fig.~\ref{fig_sed}.

We can directly compare the ring radius with the near-IR reverberation
radius $\Rlag$, which is the light-crossing distance over the time lag
of the K-band flux variation relative to the optical flux variation
(e.g. \citealt{Glass04}; \citealt{Suganuma06}). The time-lag radii
$\Rlag$ of a sample of galaxies have been shown to be approximately
proportional to $L^{1/2}$ (\citealt{Suganuma06}; their fit is shown as
a dotted line in Fig.~\ref{fig_radius_pc}), and are likely to be
probing the dust sublimation radius in each object.
Figure~\ref{fig_radius_pc} shows that the interferometric ring radii
also scale approximately with $L^{1/2}$, and they are of the same
order as the reverberation radii for a given luminosity. This is what
we reported in Paper~I, but now we confirm this with a larger
sample. This further supports that the KI observations indeed
partially resolve the dust sublimation region.

We note that all of our KI targets have similarly high visibilities,
thus the inferred angular sizes are relatively similar
(Table~\ref{tab_obs}).  This is what we expect for the $L^{1/2}$
scaling, since the angular sizes of the dust sublimation radius scale
only with the apparent magnitudes of the illuminating source (for $z
\ll 1$ cases; e.g. \citealt{Kishimoto07}), which are within a
relatively small range.  Conversely, the similarly high visibilities
observed for the KI targets imply that the physical size of the
corresponding, partially resolved structure scales roughly with
$L^{1/2}$.

\subsection{Inner radial structure}

Furthermore, we confirm the tendency already seen in Paper~I for
$\Rring$ to be either roughly equal to or slightly larger than $\Rlag$
for a given luminosity.  As we  described in Paper~I, this can
be understood in the picture where $\Rlag$ represents a radius very
close to the inner boundary of dust distribution, while $\Rring$
corresponds to a brightness-weighted effective radius.  The former is
based on the fact that the reverberation technique generally places
weight on the small responding radii when the lag is determined from
the peak in the cross-correlation function (e.g.  sect.~3 in
\citealt{Koratkar91} and references therein). It is also based
on the fact that the $\Rlag$ fit actually is very close to the dust
sublimation radius at the large-grain (black-body) limit, which is the
smallest possible sublimation radius for a given sublimation
temperature (this temperature being constrained by the observed color
temperature; \citealt{Kishimoto07}; Fig.~\ref{fig_sed}).  In this
picture, $\Rring$ close to $\Rlag$ means a steep and compact mass
distribution where most dust sits close to the innermost radius. On
the other hand, if $\Rring$ is observed to be much larger than
$\Rlag$, the dust distribution is much shallower and more extended.

In this case, we can use the ratio of $\Rring$ to the $\Rlag$ fit as
an approximate probe for the radial distribution of the innermost,
likely accreting (e.g. \citealt{Krolik88}), dusty material.  In
practice, the real inner boundary radius of the dust distribution is
probably still slightly smaller than $\Rlag$. In addition, the
measured individual $\Rlag$ has on average an uncertainty of the order
of the scatter from the fit (Fig.~30 in \citealt{Suganuma06}).  Taking
the ratio with respect to the $\Rlag$ {\it fit}, however, we can take
out the relatively robust $L^{1/2}$ dependence from the
interferometric ring radii, quite insensitively to the uncertainty in
the reverberation measurements.  The resulting ratio, at least in a
relative sense over the sample, has a physically interpretable meaning
described above, and can be used to explore any possible relations
between the inner radial structure and the property of the central
engine in each AGN. In this way, we can exploit the power of the
interferometry, which can easily distinguish a size difference of a
factor of few, as demonstrated e.g. in \cite{Kishimoto07}.

\begin{figure}
\centering \includegraphics[width=8cm]{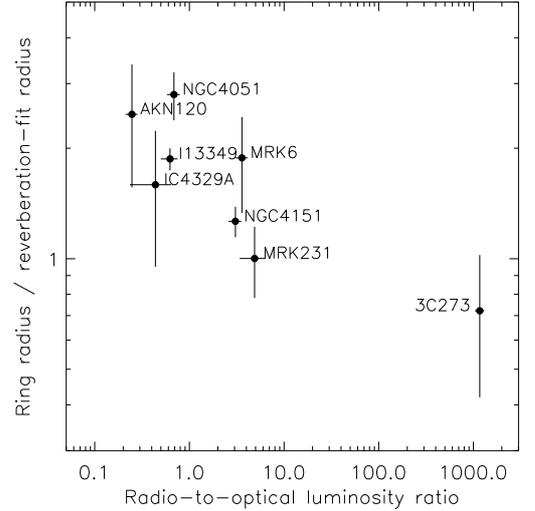}
\caption{\small Ratio of $\Rring$ to $\Rlag$ fit plotted against the ratio of
  luminosity $L_{\nu}$ at 5.0~GHz to that at 0.55 $\mu$m.  The former
  is calculated from the radio flux listed in Table~\ref{tab_obs},
  assuming $S_{\nu} \propto \nu^{-0.8}$ for extrapolation and
  K-correction. }
\label{fig_radstr_RL}
\end{figure}

\subsection{Radio connection}

Intriguingly, objects with the ratio $\Rring/\Rlag$ close to unity,
i.e.  corresponding to the steepest and most compact distribution, are
those that are known to show linear radio jet structures (NGC4151,
\citealt{Mundell03}; Mrk231, \citealt{Ulvestad98}, known as compact
symmetric object; 3C273).  Mrk6 is also known to exhibit a linear jet
structure (e.g.  \citealt{Kukula96}), but our $\Rring$ measurement has
a relatively large uncertainty.  Figure~\ref{fig_radstr_RL} shows the
$\Rring/\Rlag$ fit ratio plotted against the radio-to-optical
luminosity ratio.  Here we might have started to see evidence of a
correlation, suggesting that jet-launching objects possess a very
steep and compact distribution of inner accreting material.  We regard
this possible correlation as tentative, since the sample size is small
and the statistical significance is still not very high (e.g. the
Spearman's rank correlation analysis gives 91\% confidence, or
1.7$\sigma$).

One possible remaining ambiguity concerns the UV luminosity $L$.  Here
we have adopted the scaled optical luminosity for $L$. It might be
better to use an observed UV luminosity, perhaps averaged over a long
timescale if sublimation radius is not changing quickly.  However,
uncertainties from foreground absorption in the UV can be
significant. In addition, the observed $L^{1/2}$ scaling of $\Rlag$
seems to imply that the optical luminosity is at least approximately
proportional to the (dust-heating) UV luminosity.  We will have to
see whether further study of the most relevant luminosity strengthens
or weakens the correlation.

Another source of ambiguity is the inclination, although we focus
intentionally on Type 1 objects, which should largely avoid this
problem. In the simplest picture of an inclined circular ring, the
projection correction would make the ring radii systematically larger,
but would not necessarily weaken the observed tentative
relationship. With more measurements, we hope to pursue the PA
dependency in each object as we have done for NGC4151
(sect.~\ref{sec_ngc4151}; Fig.~\ref{fig_ngc4151}).

Finally we note that the effect of the synchrotron emission
contribution to the deduced ring size is considered to be very small,
if at all present, as we discuss below.

\subsection{Synchrotron contribution and 3C273}

For radio-loud objects, the synchrotron contribution at K-band can
potentially make the apparent ring size smaller than it actually is,
just as the unresolved AD contribution does. Based on the radio to
infrared SEDs (\citealt{Sanders89}; \citealt{Elvis94}; those from
NED), the only object for which this issue is relevant in our present
sample is 3C273.  In a quiescent state, the object is known to have
only low polarization through the optical ($<\sim0.5\%$;
e.g. \citealt{Impey89}) to near-IR ($<$1\%; \citealt{Smith87};
\citealt{Sitko91}), while a flaring state is accompanied by much
higher polarization (\citealt{Courvoisier88}; \citealt{Wills89};
\citealt{Cross01}).  The additional emerging flux in the K-band in the
latter state can be attributed to synchrotron contribution
\citep{Courvoisier88}.

However, the object was in a quiescent state at the time of both the
KI and UKIRT runs, judging from the public light curve\footnotemark[7]
and the observed flux.  We therefore infer that the K-band synchrotron
contribution, if present at all, was small at the time of our
observations. Even if all the possible K-band polarized flux in a
quiescent state, $\lesssim 0.8$~mJy, originates from synchrotron
emission with an intrinsic polarization of $\sim$10\% \citep{Impey89},
the contribution is $\lesssim 8$~mJy.  Our effective radius estimate
for the dust distribution would become larger by $\lesssim$ 10\% after
the potential correction for the synchrotron component. This is well
within the error bar shown in Figs.~\ref{fig_radius_pc} and
\ref{fig_radstr_RL}.

\footnotetext[7]{http://www.aavso.org}

\section{Conclusions}

We have presented Keck interferometer measurements at 2.2~$\mu$m for 4
new Type 1 AGNs, as well as re-observations of NGC4151.  Putting these
together with the results from our previous targets and comparing
these with the optical-infrared reverberation measurements, we have
interpreted the observed high visibilities as an indication of the
partial resolution of the dust sublimation region in each object.

We have argued that we can interpret the ratio of the interferometric
effective ring radius of the inner dusty region to the reverberation
$L^{1/2}$-fit radius as an approximate probe of the steepness/flatness
of the inner radial structure.  It has been tentatively shown that
this inner structure might possibly be related to the radio-loudness
of the central engine.

The possible radio connection at least illustrates the important
potential of this new exploration of the inner region.  We are still
limited in terms of the proximity to the central engine, since we are
using the emission from dust grains which cannot survive at distances
too close to the central UV source.  However, the dust distribution in
or around the sublimation region is still a direct probe of the
physical structure of the accreting material at the best possible
spatial resolution available so far.  We plan to further explore the
region with more interferometric measurements.

\begin{acknowledgements}



  The data presented herein were obtained at the W.M. Keck
  Observatory, which is operated as a scientific partnership among the
  California Institute of Technology, the University of California and
  the National Aeronautics and Space Administration (NASA). The
  Observatory was made possible by the generous financial support of
  the W.M. Keck Foundation.  The Keck Interferometer is funded by NASA
  as part of its Exoplanet Exploration program.  The United Kingdom
  Infrared Telescope is operated by the Joint Astronomy Centre on
  behalf of the Science and Technology Facilities Council of the U.K.
  We thank N.~Teamo and J.~C.~Pelle for kindly providing the
  pre-imaging data.  This work has made use of services produced by
  the NASA Exoplanet Science Institute at the California Institute of
  Technology.  This research has also made use of the NASA/IPAC
  Extragalactic Database (NED) which is operated by the Jet Propulsion
  Laboratory, California Institute of Technology, under contract with
  NASA. SH acknowledges support by Deutsche Forschungsgemeinschaft
  (DFG) in the framework of a research fellowship
  (``Auslandsstipendium'').

\end{acknowledgements}


\end{document}